\documentclass[aps,prl,preprint,groupedaddress]{revtex4}
\usepackage{amsmath,amssymb,amsfonts} 
\usepackage{color}                    
\usepackage{hyperref}                 
\usepackage{subfigure}
\usepackage{epsfig}

\begin{document}
\title{Simulation Study of the Beam Calorimeter in the GLD Configuration for
the Next Generation Linear Collider}
\author{S.-X.~Huang}
\affiliation{Department of Physics, National Taiwan University, 
Taipei 10617, Taiwan, R.O.C.}
\author{M.-C.~Chang}
\affiliation{Department of Physics, Fu Jen Catholic University, 
Taipei 24205, Taiwan, R.O.C.} 
\author{M.-Z.~Wang}
\affiliation{Department of Physics, Institute of Astrophysics and
Leung Center for Cosmology and Particle Astrophysics, National Taiwan
University, Taipei 10617, Taiwan, R.O.C.}

\date{\today}

\begin{abstract}

The beam calorimeter, located in the very 
forward and backward region of the next generation $e^+e^-$ linear accelerator, 
will be an important apparatus to search for or identify 
new particles beyond Standard Model. One of its key design issues is to 
estimate the radiation dosage due to beam related backgrounds. We used the
geant-based package, Jupiter, for the proposed GLD detector in this
study. The dosage received per year for the inner most ring will be around 
10 Mrad. An algorithm for electron identification with beam calorimeter
under nominal background condition is developed and is used for the
feasibility study of smuon search. The result is satisfactory.

\end{abstract}

\maketitle

\section{Introduction}
The International Linear Collider (ILC)~\cite{ilc} is a proposed next generation
 electron-positron linear collider. 
It is expected to operate in an initial 
center-of-mass energy of 200~GeV - 500~GeV, with a peak luminosity 
of $2\times10^{34}$~$\rm {cm^{-2}s^{-1}}$.
From previous study~\cite{slac371}\cite{kek91-2}, the beamstrahlung contributes the dominant
  energy deposition in the small angle region along the beam direction.
 When the high energy bunches of electron and positron collide at interaction point (IP), due to the strong electromagnetic field of the oppsite bunches, the beam particles would be deflected and the beam cross section would be reduced. The bent trajectory of electrons and positrons from both bunches will emit forward photons, which are called beamstrahlung photons. Simulation programs, named \textbf{CAIN}~\cite{cain21e} and \textbf{Guinea-Pig}~\cite{guineapig}, have been developed to investigate these effects.

The energy loss from the beamstrahlung photons is about 2\% of beam energy 
on the average (Nominal beam set)~\cite{beamset},
 and this is the largest energy loss among the machine backgrounds.
 The beamstrahlung photons are very close to the beam line ($\theta <$ 0.5 mrad). Some of beamstrahlung photons would
 interact with beam particles from opposite direction and
 produce electron-positron pairs, which are called incoherent pairs.
 Incoherent pairs are the main background source to the detector in
 the very forward and backward region.
 We use the full geant simulation package, Jupiter~\cite{jupiter}, 
  for the proposed GLD~\cite{glddod} at ILC 
  to study the beam calorimeter (BCAL) performance.
Several studies in other detector models can be found elsewhere~\cite{lohmann}.

\begin{figure}[h!]
\centering
\includegraphics[width=3.5in]{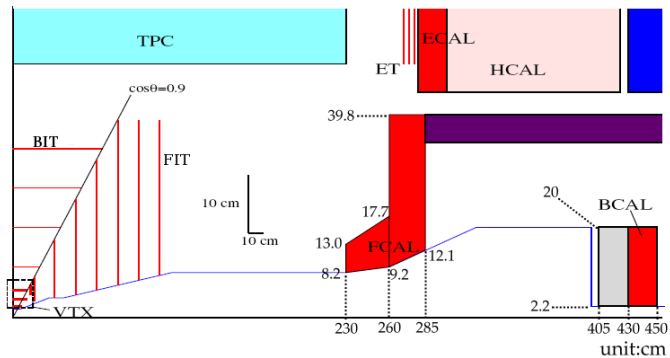}
\caption{Schematic view of the GLD detector in the forward region.}
\label{fig:fcal01}
\end{figure}

The shape of BCAL is a cylinder which consists of 33 layers of tungsten sandwiched silicon
with a thickness of 3.5~mm for tungsten and 0.3~mm for silicon.
 It is placed at 430~cm to 450~cm from the IP (Figure~\ref{fig:fcal01}), and the corresponding polar angle is 5~mrad to 50~mrad. 
BCAL can be used as a fast beam diagnostic device to analyze
the activities at small angle which is crucial for 
the new particle search. For example, BCAL is very important for searching
for the stau pair production~\cite{stau}. The mass of stau particle
might be able to explain the amount of dark matter in the universe.

   In this paper, we choose the dominant background process, 
namely beamstrahlung incoherent pair, to give an order of magnitude 
estimation of the expected dosage received by BCAL per year 
under the nominal beam condition.
This information is very important for the design limitation of BCAL. 
We also develop an algorithm for the BCAL electron identification 
with incoherent pairs as the principle background source. 
Its performance is studied by a search for smouon which is closely related
to the study reported for stau.

\section{Simulation Parameters}

Parameters in our simulation are according to the GLD 
configuration - 3~Tesla magnetic field, 500~GeV center-of-mass energy of beams,
 and 2~mrad beam crossing angle. Each layer of BCAL is sliced into 18 segments,
 from the inner radius (2.2~cm) to the outer radius (20.0~cm) along the radial direction. The size of azimuthal segments in each layer is 1~cm. For Nominal beam set, there are 2,820 bunches in each train (5~Hz) and the time interval of adjacent bunch is 307~ns,
 which is manageable with conventional pipeline readout.
BCAL has about 20,000 channels, and it will produce 40~kB data for a bunch crossing if each channel needs 2 bytes in digitalization.
 The peak data rate will be 130~GB/s, which is practically impossible.
 We assume that some zero suppression techniques and low level
 trigger scheme will be implemented such that interesting events could be
saved.

\begin{figure}[h!]
\centering
\subfigure{
\label{fig:edeplayer}
\includegraphics[width=2in]{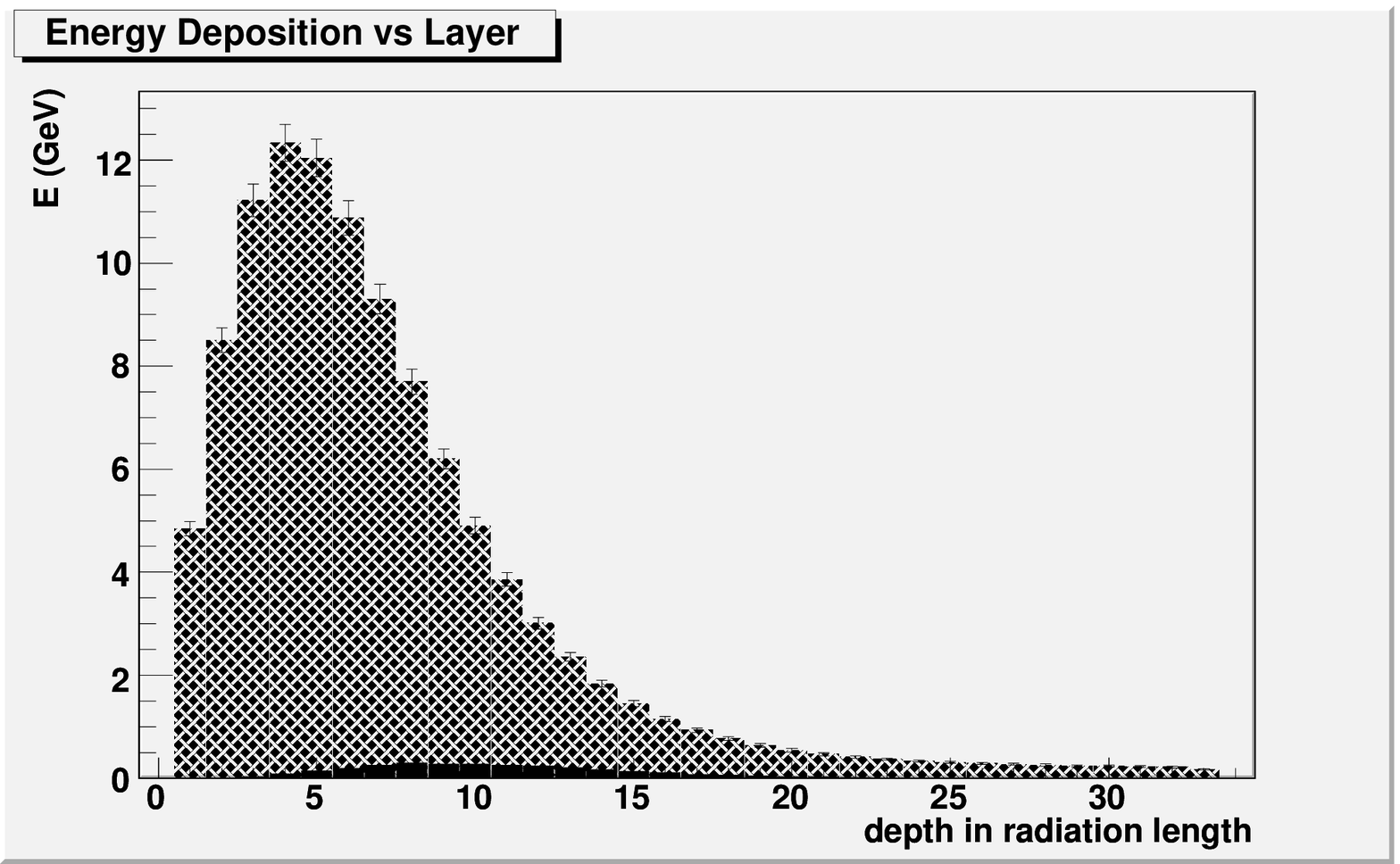}
}
\subfigure{
\label{fig:bunch003}
\includegraphics[width=1.2in,height=1.2in]{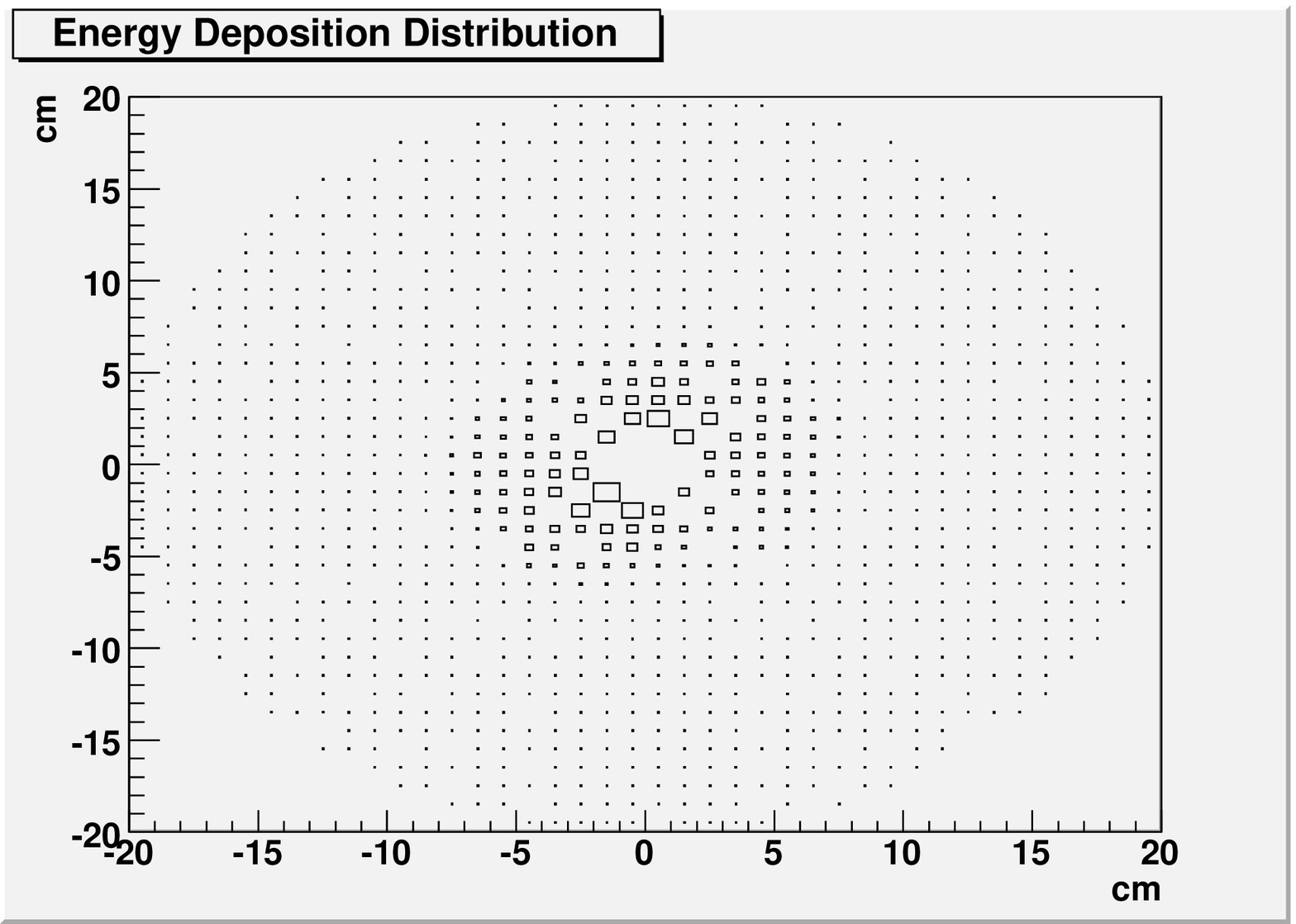}
}
\caption{Energy deposition in the BCAL. Left: Comparison of energy deposition in the beam calorimeter between incoherent pairs per bunch (the shadow region) and a single 250~GeV electron (the black region). Right: Energy distribution of the fifth layer of the BCAL from incoherent pairs.}
\label{fig:edep}
\end{figure}

One hundred bunch crossings (BX) of incoherent pairs in Nominal beam set are generated with CAIN~\cite{cainp500} and passed to Jupiter.
For a bunch crossing, both forward and backward side of BCALs would recieve energy about 8.5~TeV from incoherent pairs. The energy deposition in BCAL for  both signal and incoherent pairs in each layer is shown in Figure~\ref{fig:edep}. We can see that the energy from incoherent pairs is much higher comparing to a 250~GeV electron. 

The energy deposition (without calibration) of HighLum~\cite{beamset} beam
parameter set was shown for comparsion in Table~\ref{tb:obps}. 
The luminosity of HighLum beam set is $4.9\times10^{34}$~$\rm
{cm^{-2}s^{-1}}$. It produces much more background, and only 50~BX were used in the simulation due to long
computation time. The 100~BX should have twice events comparing with 50~BX
data. 
Although this is just for a quick check, the mean values listed in Table~\ref{tb:obps} wouldn't
vary too much which is good enough for our study.

\begin{table}[h!]
\centering
\caption{Energy deposition of two different beam parameter sets}
\label{tb:obps}
\begin{tabular}{|l|l|l|}
\hline
BeamSet&Forward E (GeV)&Backward E (GeV)\\
\hline
Nominal&108.3&108.2\\
HighLum&603.3&602.2\\
\hline
\end{tabular}
\end{table}

\section{Radiation Dose}
\begin{figure}[h!]
\centering
\subfigure[Nominal]{
\label{fig:erz1}
\includegraphics[width=1.6in]{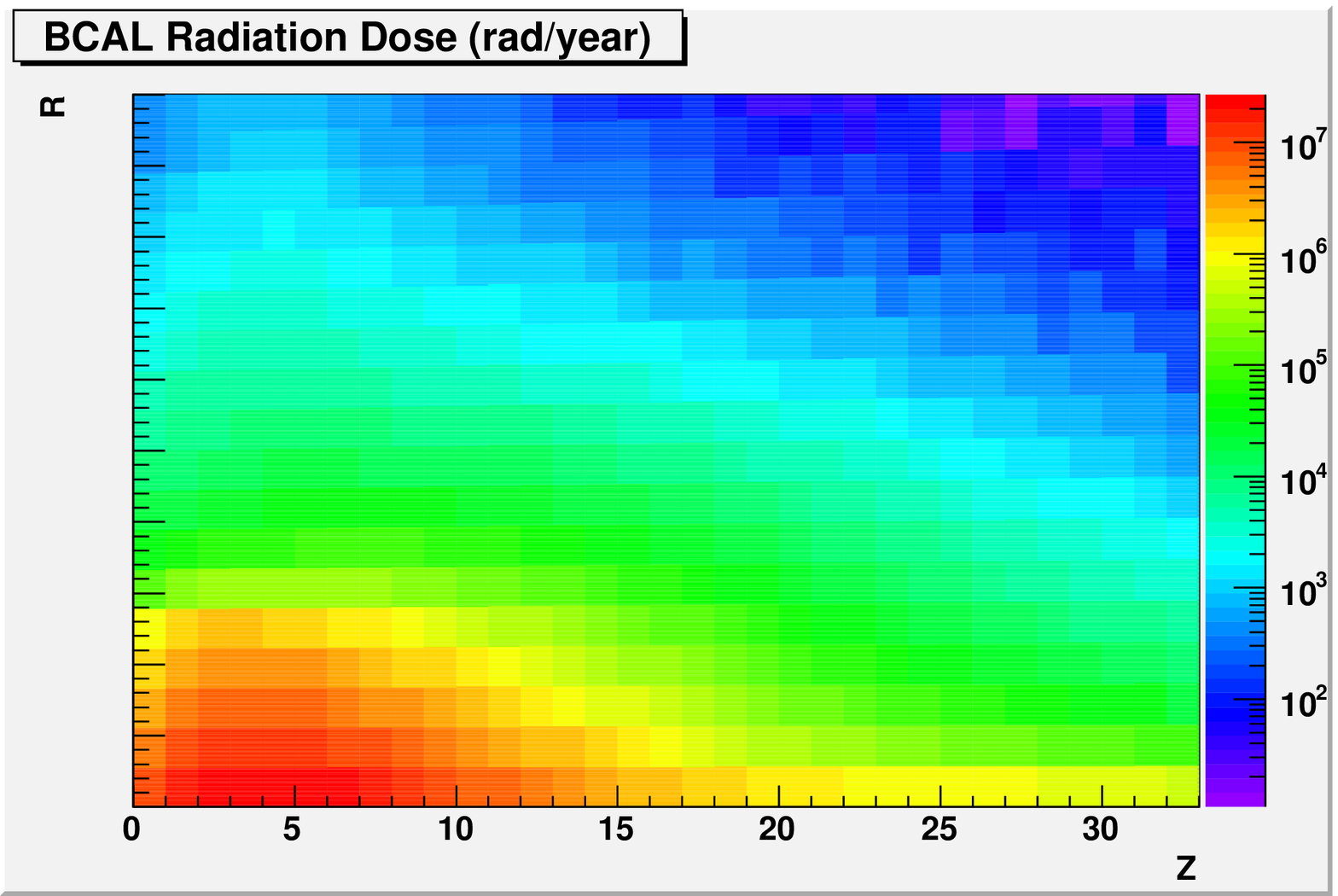}
}
\subfigure[HighLum]{
\label{fig:erz2}
\includegraphics[width=1.6in]{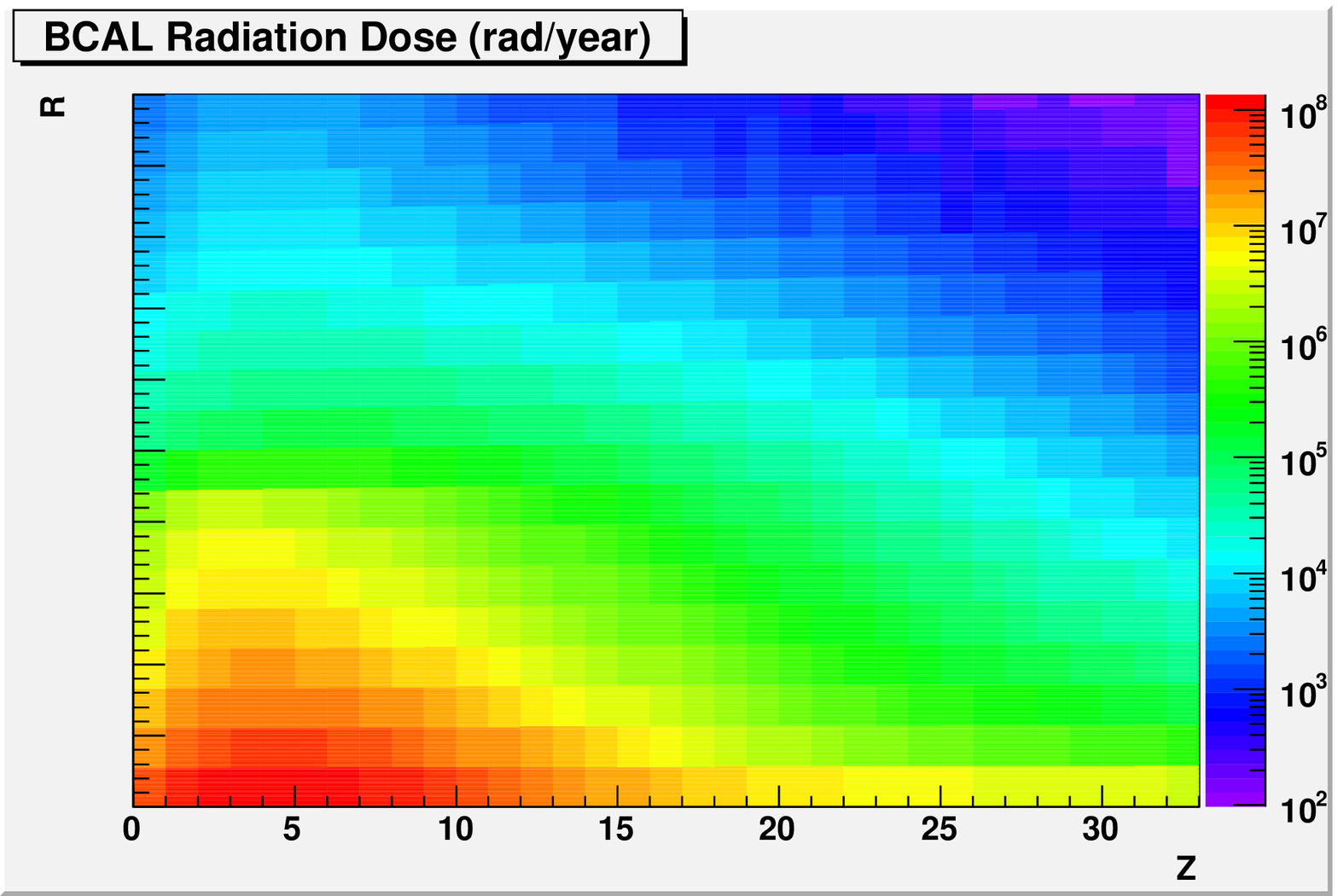}
}
\caption{Radiation dose of beam calorimeter from incoherent pairs.}
\label{fig:erz}
\end{figure}

The radiation tolerance of silicon detector is an important issue. 
A conventional silicon detector can be operated up to the radiation dose of several Mrad~\cite{rad900}. For BCAL, if it received large energy deposition, meaning that the serious radiation damage happened.
 Figure~\ref{fig:erz} shows
the radiation dosage from incoherent pairs deposited in BCAL in the $\rm z-r$
plane from Jupiter simulation. We could see extremely high radiation dosage ($\sim$10~Mrad/year) at the small angle rings. For HighLum beam set, the radiation
 dosage can be as high as 100~Mrad per year. 
Radiation hard sensors are needed for this purpose~\cite{radhard}.

\section{Electron Identification}

When the background particles of a bunch crossing hit BCAL,
 they will leave energy in the BCAL cells.
 To check the electron identification efficiency under the beamstrahlung backgrounds, 250~GeV electrons are randomly generated to BCAL in Jupiter and passed to the "electron finder".
The electron finding relies on a clustering algorithm
and a comparison with the electromagnetic cascade shower profile.
However, assuming no special treatment for background suppression,
the identification of electrons below 20 mrad is totally failed because of
the severe background.
 It can be easily understood in Figure~\ref{fig:edep}, because most of the energy deposition from backgrounds is in the small angle rings.

The basic idea of suppressing the effects from incoherent pairs is to substract mean
deposition energy of bunch crossings in each cell. The following procedure was used to get the electron identification efficiency with incoherent pairs:
\begin{enumerate}
\item
We generate 100~BX Nominal and 50~BX HighLum incoherent pairs from CAIN. The generated events were submited to the full geometry simulator, Jupiter. 
\item
We calculate the mean and variance of deposited energy in each cell of the beam calorimeter, and randomly choose a bunch data. Then,
we superimpose randomly generated electron signals onto this data set.
\item
Finally, we substract the mean deposited energy of each cell and set the $E_{fired}$=1$\sigma$ for each cell. $E_{fired}$ is a parameter of electron finder. The cell will not be included in the electron finder if energy deposition is
below $E_{fired}$. 
\end{enumerate}

\begin{figure}[h!]
\centering
\subfigure[Nominal]{
\label{fig:eff}
\includegraphics[width=1.6in]{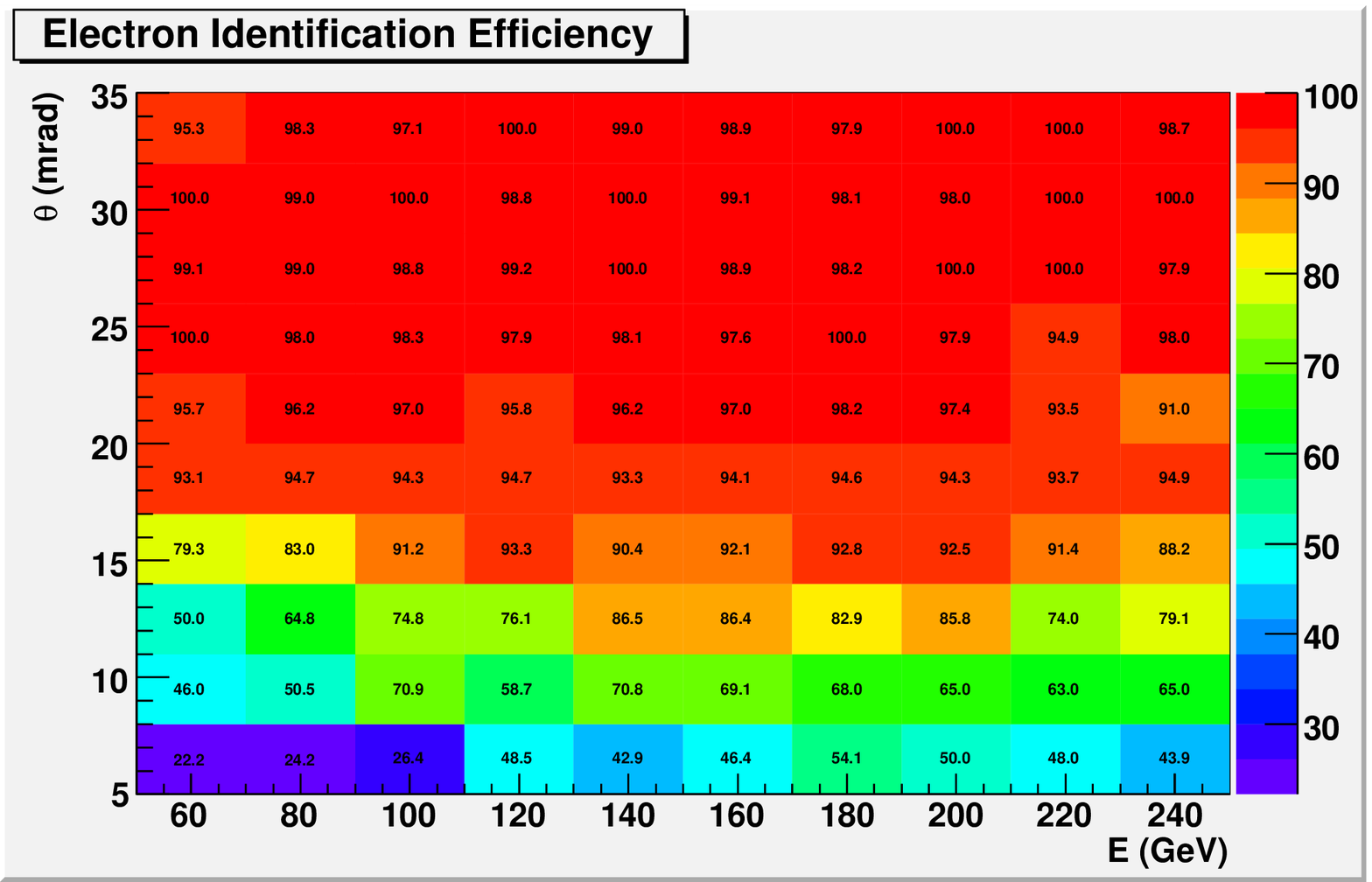}
}
\subfigure[HighLum]{
\label{fig:heff}
\includegraphics[width=1.6in]{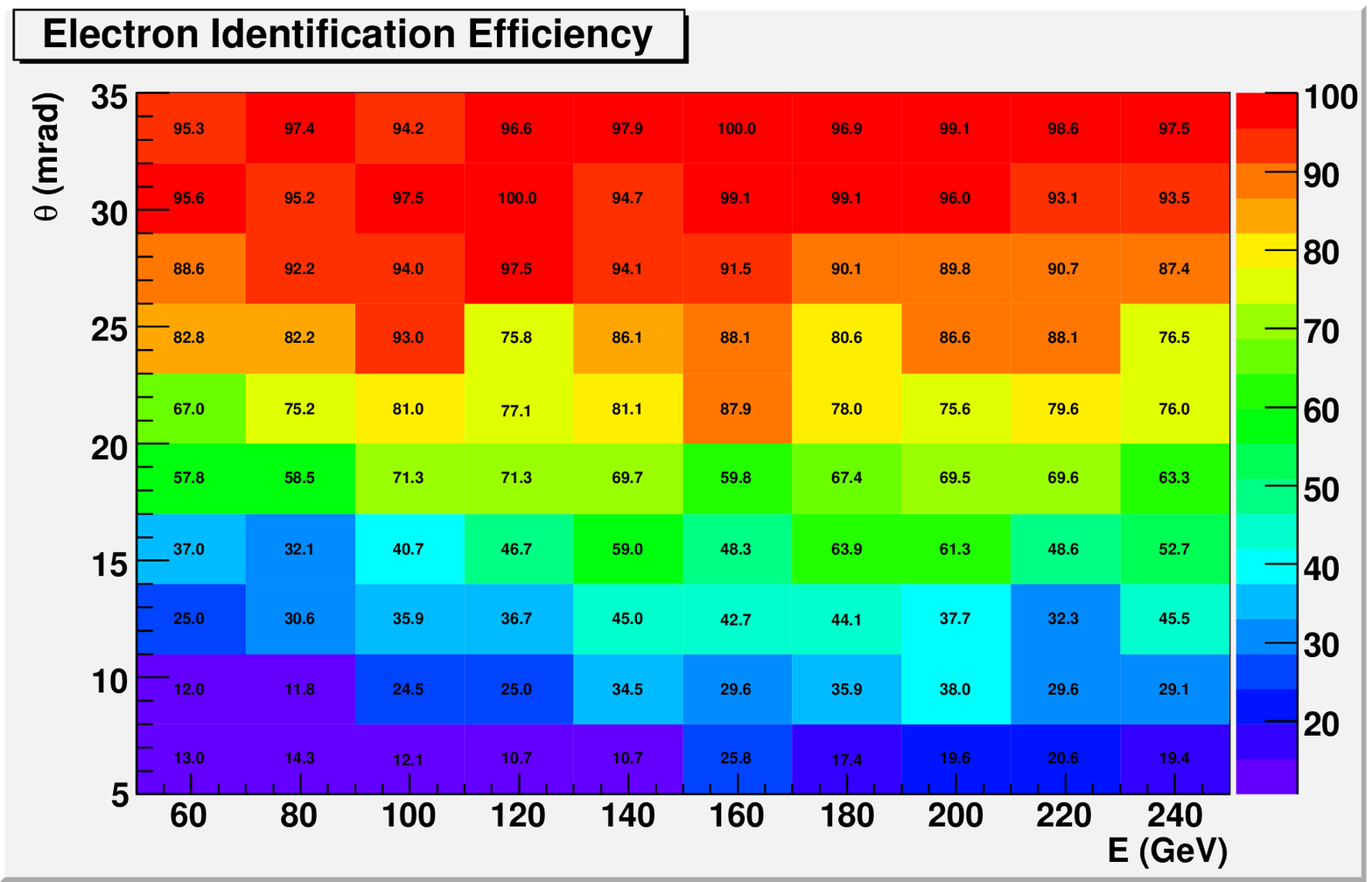}
}
\caption{Electron identification efficiency (\%) of a) Nominal beam set and b)
HighLum beam set.}
\label{fig:aeff}
\end{figure}
\begin{figure}[h!]
\centering
\includegraphics[width=3.5in]{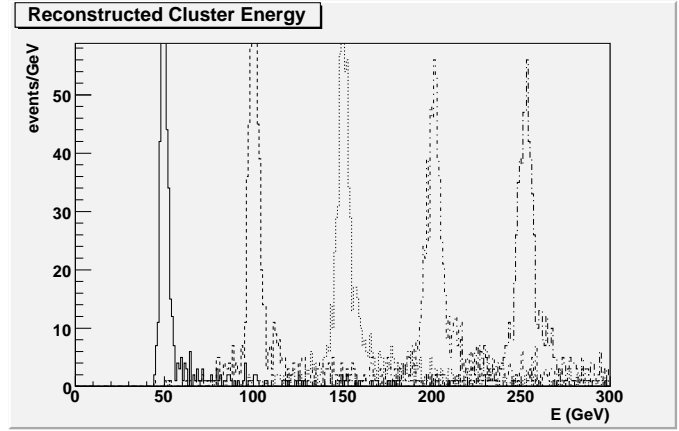}
\caption{Reconstructed cluster energies (50~GeV, 100~GeV, 150~GeV, 200~GeV, and 250~GeV) in Nominal beam set.}
\label{fig:eres}
\end{figure}

The electron identification efficiency is shown in Figure~\ref{fig:aeff}.
Electron identification efficiency depends on incident electron energy and also the energy
deposition of incoherent pairs. The smaller polar angle region shows worse efficiency.
 As one can expected, the beam parameter set with the larger beamstrahlung effect showed poorer electron identification efficiency. 
However, eletron identification efficiency is generally independent to the incident energy at lower background region. Energy resolution is also affected by the background, as shown in Figure~\ref{fig:eres}.

\section{Smuon Search}
The masses of smuon, $\tilde{\mu}_R$, and the neutralino, $\tilde{\chi}^0_1$,
 can be evaluated by analyzing the energy spectrum of the final state leptons $\mu^+$ and $\mu^-$.
   The right-handed smuon decays isotropically according to the decay mode
\[
e^+e^-\rightarrow \tilde{\mu}^+_R\tilde{\mu}^-_R
\rightarrow
{\mu}^+\tilde{\chi}^0_1{\mu}^-\tilde{\chi}^0_1
\]
where $\tilde{\mu}$ is supersymmetric partner of the muon, and it will decay into one standard model particle and one
sparticle - muon and neutralino.
It is believed that neutralino is the least massive supersymmetric particle (LSP).
The neutralino interacts via weak force only, 
and it will penetrate through our detector without being detected.
Since the neutralino cannot be detected directly, the experimental signature is 
the muon pair only. Several standard model backgrounds are considered to be important.
Those are $e^+e^-\rightarrow ZZ\rightarrow \mu^+\mu^-\nu\bar{\nu}$, 
$e^+e^-\rightarrow W^+W^-\rightarrow \mu^+\nu_\mu\mu^-\bar{\nu_\mu}$,
 $e^+e^-\rightarrow \tau^+\tau^-\rightarrow \mu^+\nu_\mu\bar{\nu_\tau}\mu^-\bar{\nu_\mu}\nu_\tau$ and $e^+e^-\rightarrow e^+e^-\mu^+\mu^-$.

Ten thousand events are generated with SUSYGEN3~\cite{susygen}, and the total cross section is 86.6~fb at 500~GeV.
The electron beams are 80\% right-handed polarized.
Assuming R-parity conservation, five parameters are used in the event
generator according to SPS1a~\cite{sps1a}
\[ m_0=100~GeV,m_{1/2}=250~GeV \]
\[ A_0=-100~GeV,tan\beta=10,\mu>0 \]
where $m_0$ is the universal scalar constant,
 $m_{1/2}$ is the universal gaugino mass,
 $A_0$ is the universal trilinear term,
 $tan\beta=v_2/v_1$ is the ratio of the vaccum expectation values of two Higgs doublets,
and $\mu$ represents the supersymmetry conserving higgsino mixing term.
The mass of neutralino is 98.0~GeV and the mass of right-handed smuon is 144.7~GeV.
In the smuon decay, the simple two-body kinematics results in a flat distribution of the observed muons.

One million two photon background $e^+e^-\rightarrow e^+e^-\mu^+\mu^-$ events are generated with
GRC4F~\cite{grc4f}, a four-fermions event generator.
Several selections are applied at the generator level to save
 computing time to generate useful background events:
\begin{enumerate}
\item
$E_{\mu^{\pm}}^{min}$=10 GeV
\item
$E_{\mu^{\pm}}^{max}$=130 GeV
\item
$7^{\circ}<\theta_{\mu^{\pm}}<173^{\circ}$ (Detector acceptance for muon
identification)
\end{enumerate}

Other backgrounds listed in Table~\ref{tb:bgcut} are generated with
PANDORA\_PYTHIA~\cite{pandora} which includes 
the ISR and beamstrahlung effects. The electron beams are set with 80\% right-handed polarization.
The detector acceptance for electron identification without BCAL can be down to 35 mrad.

The event selection criteria are listed below: 
\begin{enumerate}
\item
Muon pair acoplanarity, $|\phi_{aco}|<2.9 (166^\circ)$. 
Acoplanarity is the $\phi$ difference of $\mu^+\mu^-$ pair.
\item
Transverse momentum of $\mu^\pm$ ($P_T$)$ > 10 \rm GeV $.
\end{enumerate}
The events which don't fit above criteria are regarded as backgrounds.
To suppress two photon background, there is a special veto 
that if an electron or a positron is found by the main detector
and the total missing $P_T$ of the event is less than 2 GeV, that event will
be identified as two photon background and will be vetoed.

Table \ref{tb:bgcut} shows different physical cross sections before and after
   the event selection.

\begin{table}[h!]
\centering
\caption[List of background processes]
{Cross section before and after event selection. Electron beams are set with 80\% right-handed polarization.
 Note that these selections are applied on four momentum directly and detector effect is not included.}
\label{tb:bgcut}
\begin{tabular}{lrrr}
\hline
 Process&$\sigma$ (fb)&Efficiency&Visible $\sigma$ (fb)\\
\hline
\begin{tabular}{l}$e^+e^-\rightarrow\tilde{\mu}^+\tilde{\mu}^-$\\
\hspace{8pt}$\rightarrow \mu^+\tilde{\chi}^0_1\mu^-\tilde{\chi}^0_1$
\end{tabular}&86.6&0.798&69.15\\
\hline
$e^+e^-\rightarrow e^+e^-\mu^+\mu^-$&$53.09\cdot 10^3$&$8.24\cdot 10^{-3}$&437\\
\hline
\begin{tabular}{l}$e^+e^-\rightarrow W^+W^-$\\
\hspace{8pt}$\rightarrow \mu^+\nu_\mu\mu^-\bar{\nu_\mu}$\end{tabular}
&17.29&$8.91\cdot 10^{-2}$&1.54 \\
\hline
\begin{tabular}{l}$e^+e^-\rightarrow ZZ  $\\
\hspace{8pt}$\rightarrow \mu^+\mu^-\nu\bar{\nu}$\end{tabular}
&4.70&0.13&0.62 \\
\hline
$e^+e^-\rightarrow \mu^+\mu^-$&1213.29&$< 10^{-5}$ &$< 0.012$\\
\hline
\begin{tabular}{l}$e^+e^-\rightarrow \tau^+\tau^- $ \\
\hspace{8pt}$\rightarrow\mu^+\nu_\mu\bar{\nu_\tau}\mu^-\bar{\nu_\mu}\nu_\tau$\end{tabular}
&41.69&$< 3\cdot 10^{-4}$ &$< 0.013$\\
\hline
\end{tabular}
\end{table}

\section{Two Photon Veto}
After the event selection, there are still 8,241 out of 1 million $e^+e^-\rightarrow e^+e^-\mu^+\mu^-$ events cannot be 
seperated from the signal. The visible cross section is still quite large (437~fb)
comparing to the signal, and we need to identify the escaped
 two photon events further, in order to get pure muon pair decayed from smuon.

\begin{figure}[h!]
\centering
\includegraphics[width=3.5in]{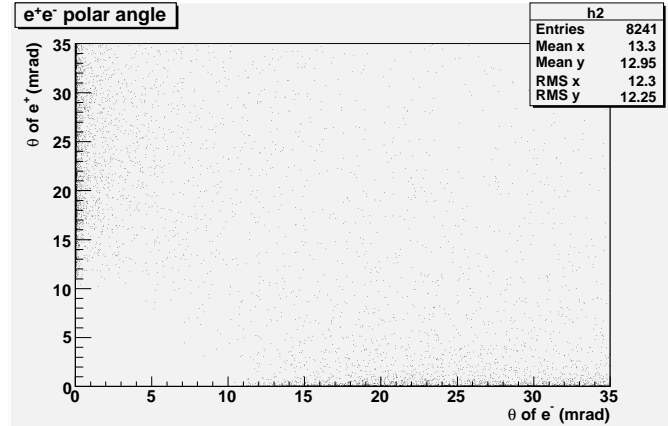}
\caption{Polar angle distribution of $e^\pm$ after the selections are applied.}
\label{fig:etptcut}
\end{figure}

In Figure~\ref{fig:etptcut}, it shows that
the polar angle of either electron or poistron of the escaped two photon events
 falls in the region between 5~mrad to 35~mrad. 
The event is identified as two photon event if
either muon pairs and one of the electron pairs are detected at the same
time, and therefore the hermeticity of the beam calorimeter down to 5~mrad 
 is just enough for our purpose.

\begin{figure}[h!]
\centering
\subfigure{
\label{fig:bgsigadd2}
\includegraphics[width=1.6in]{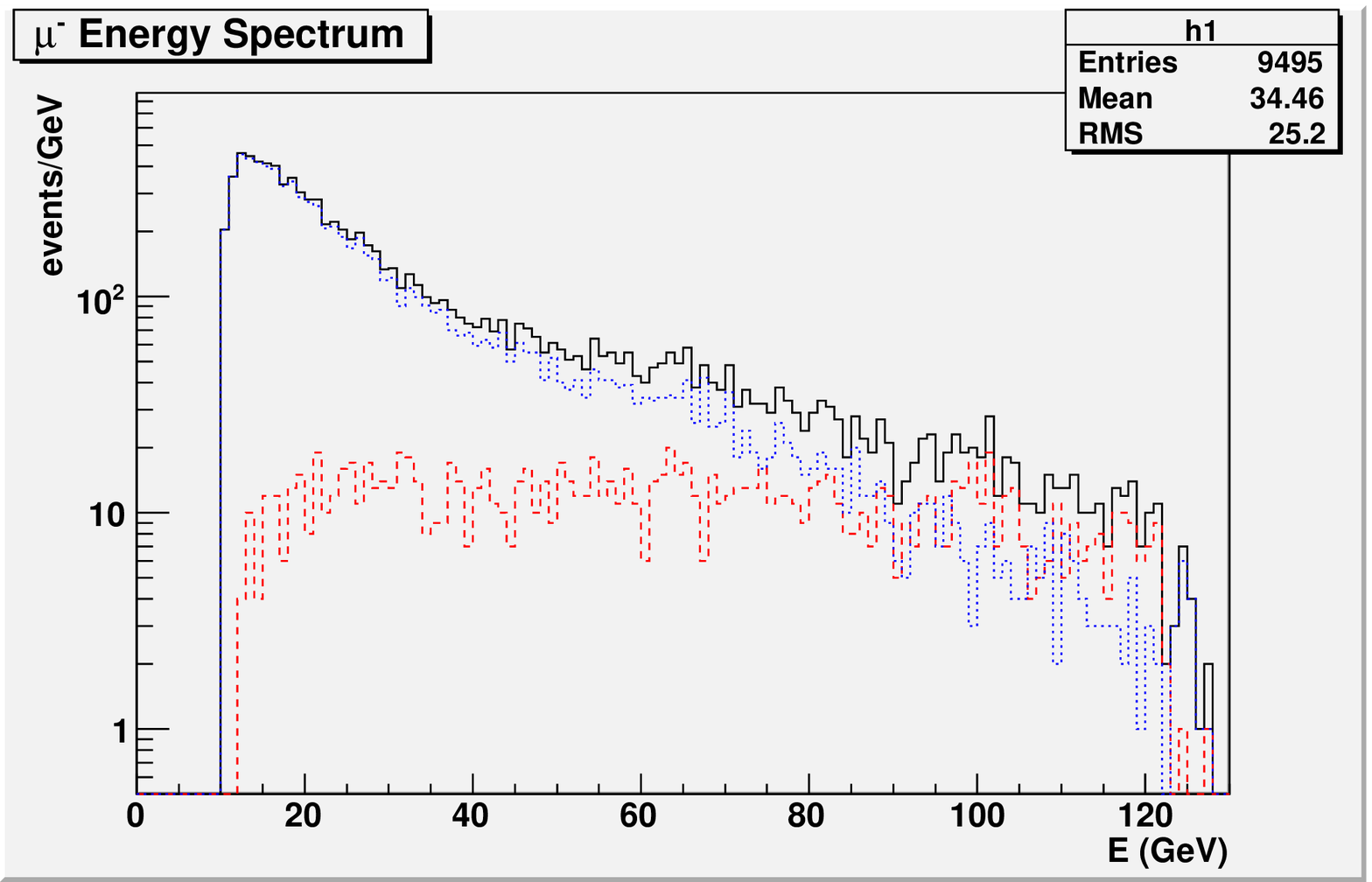}
}
\subfigure{
\label{fig:bgsigadd}
\includegraphics[width=1.6in]{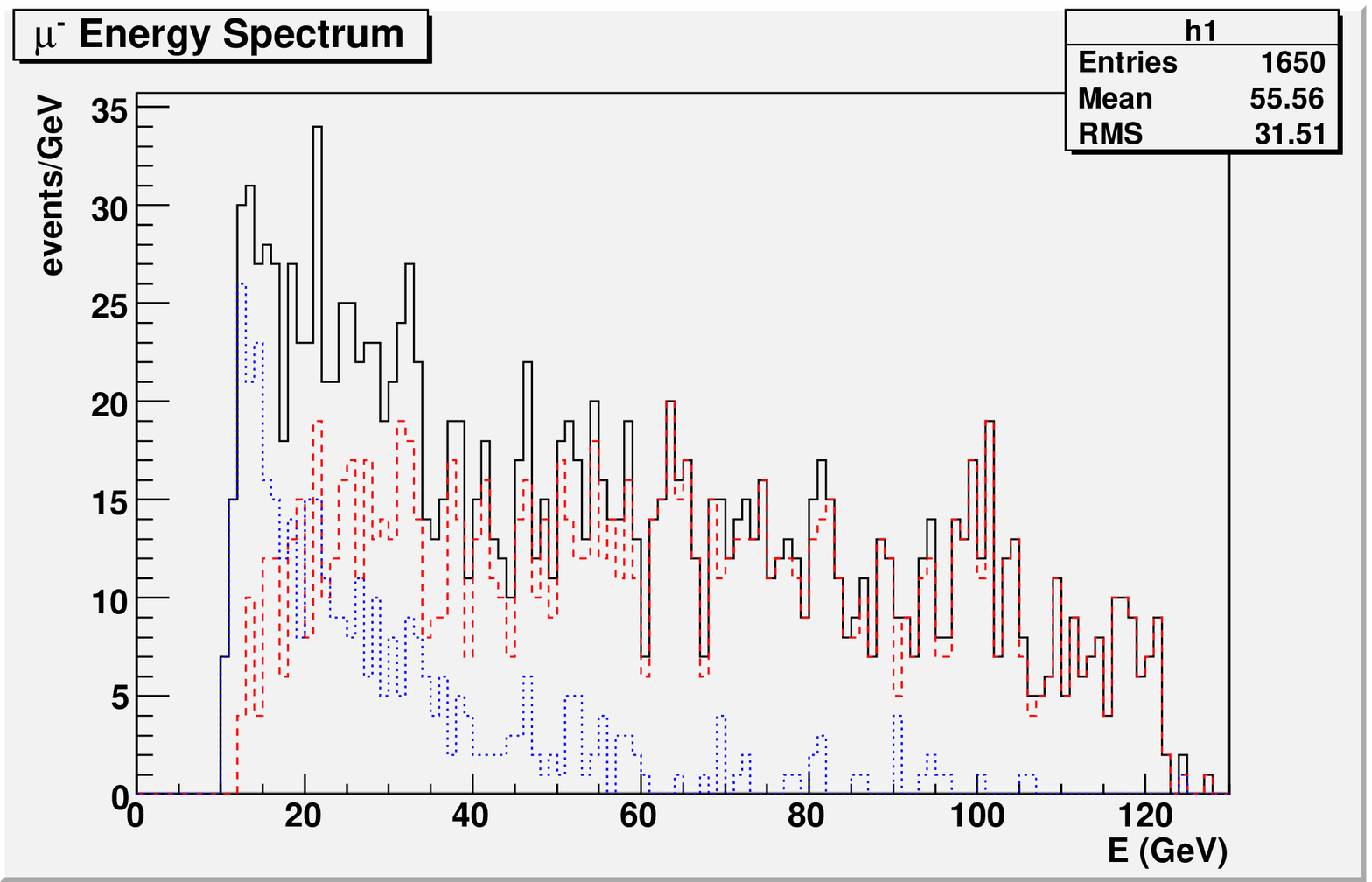}
}
\caption{Muon energy spectrum of the signal and background. The left figure is before two photon veto, and the
right one is after two photon veto from simulation.
Blue (dotted) line is two photon background,
Red (dashed) line is the signal, and black line is the superposition of signal and background.
The MC data corrosponds to 18.8~$\rm fb^{-1}$}.
\label{fig:bgsig}
\end{figure}

There are 8,241 events selected to do full simulation. With the consideration of
 beamstrahlung effect, there are 7,845 (95.2\%) events being identified. 
 The visible cross section is then eliminated down to 21~fb, as shown in
Figure~\ref{fig:bgsig}.

\section{Conclusion}
Constructing the beam calorimeter placed at a very small polar angle is a
challenging technology. The major radiation source is from secondaries of beamstrahlung. 
It leaves about 8.5~TeV to each beam calorimeter for a bunch crossing.
The radiation dose of the detector varies with the rings and the
depth. At low background region, the dosage per year will be in the order of
krad. At very low angle rings, the dosage per year will be as high as 10~Mrad.
It is the maximum radiation tolerance of silicon sensor. 
To design such detector, radiation hard sensor is prefered.
Some people consider radiation hard diamond sensor sandwiches 
instead of silicon sensor in the sampling calorimeters~\cite{radhard}.  Beside
radiation damage, electron identification is also challenging due to large
energy deposition from incoherent pairs.

Electron identificaiton efficiency generally depends on the severeness of background.
It depends on the beam parameter set and the region in the beam calorimeter.
In this study, beam calorimeter plays an important role in
the search of the supersymmetric particle - smuon $\tilde{\mu}_R$ to reject
major background events.
With the help of beam calorimeter,  
all of $e^+e^-\rightarrow e^+e^-\mu^+\mu^-$ can be identified ideally.
But the beamstrahlung effects would affect the veto efficiency and 
the electron finding efficiency 
would also vary with the finding method.
A simple and basic electron finding method is used in this study
to check its performance.
From the simulation results, 95\% of the two photon background
events are vetoed with the consideration of the beamstrahlung
effect using BCAL. The result is satisfactory.

\end{document}